\documentclass{aa}
\usepackage{natbib}
\usepackage{epsfig}

\bibpunct{(}{)}{;}{a}{}{,}
\usepackage{graphicx}

\newcommand{\teff}{$T_{\rm eff}$}

\newcommand{\logg}{log\,$g$}
\newcommand{\kms}{km\,s$^{-1}$}

\begin{document}
   \title{Searching for merger debris in the Galactic halo: \\
          Chemodynamical evidence based on local blue HB stars}

   \author{M. Altmann
          \inst{1}
          \and
          M. Catelan\inst{2}
          \and 
          M. Zoccali\inst{2}
          }

     \offprints{M. Altmann}

   \institute{Departamento de Astronom\'\i a, Universidad de Chile, Casilla 36-D, 
              Correo Central, Santiago, Chile\\
             \email{martin@das.uchile.cl}
              \and    
              Pontificia Universidad Cat\'olica de Chile, Departamento de 
              Astronom\'\i a y Astrof\'\i sica, 
              Av. Vicu\~{n}a Mackenna 4860, 782-0436 Macul, Santiago, Chile\\ 
             \email{mcatelan, mzoccali@astro.puc.cl}
             }

   {\date{\today}}

   \abstract{We report on the discovery of a group of local A-type blue 
   horizontal-branch (HBA) stars moving in a prograde, comet-like orbit with 
   very similar kinematics and abundances. This serendipitously discovered 
   group contains 5 or 6 local HBA stars venturing very close to the Galactic 
   centre; their [Fe/H] is around $-1.7$, and they seem to present minimum 
   scatter in at least Mg, Si, Ti, Fe, Al, and Cr abundances. This 
   {\em ``Cometary Orbit Group''} (COG) was found while 
   we were testing a new method to detect the debris associated with the 
   merger of smaller, specific protogalactic entities into our galaxy. The 
   method is primarily intended to identify field HBA stars with similar 
   kinematics and detailed, multi-species abundance patterns 
   as seen among members of a surviving remnant (e.g., $\omega$~Centauri). 
   Quite possibly, the COG is the remnant, on a highly decayed orbit, of 
   a merging event that took place in the relatively remote past (i.e., at 
   least one revolution ago).

}
\authorrunning{M. Altmann et al.}
\titlerunning{Searching for merger debris among local blue HB stars}
   \maketitle
%

\section{Introduction}
\label{intro}
Over the past decade evidence has accumulated that the Galactic halo was at least 
partially built up by smaller entities that merged with the main body of the Galaxy. 
Particularly compelling evidence in this direction has been provided by the discovery 
of the Sagittarius dwarf spheroidal (dSph) galaxy \citep{Ibata94}. More recently, 
another object penetrating our galaxy has been found, namely the Canis Major 
dwarf galaxy
\citep{Martin2004}. One object which has been repeatedly mentioned as the possible 
remnant of a merger process is $\omega$~Centauri (NGC~5139), the most massive Galactic 
globular cluster (GC). Its unique properties 
among GCs, including a large spread in metalicities \citep{Norris1975} 
and ages (\citealt{Hilker2004} and references therein), have raised suspicions that  
$\omega$~Cen may in fact be the nucleus of a former dwarf galaxy which was accreted 
into the main body of the Milky Way several Gyr ago. This hypothesis is strengthened 
by the fact that the second most massive Galactic GC, M\,54 (NGC~6715), is in fact 
associated with the Sagittarius dSph, and may even constitute its nucleus 
(e.g., \citealt{Layden2000} and references therein). 

In the present {\em Letter}, we report on the discovery of a group of stars that 
seems likely to be associated with a merger event. This group, which we will refer 
to as the {\em Cometary Orbit Group} (COG), was 
found while we were testing a new method, involving detailed orbital information 
and multi-species element abundances for A-type blue horizontal branch (HBA) stars, 
devised to identify tidal debris related to a specific surviving protogalactic 
fragment, such as $\omega$~Cen itself. 

\section{The method} 
\label{method}

Several studies have attempted to identify the debris 
of an $\omega$~Cen-related merger event among field stars 
\citep{Dinescu2002,Bekki2003,Mizutani2003}.
All of these predict the remnant moving in a retrograde manner (as $\omega$~Cen itself). 
Our primary goal, with the proposed method, is to identify local HBA stars moving 
in a retrograde orbit ($\Theta<0$ \kms) with abundance patterns that are similar 
to those seen among evolved 
red giant branch (RGB) stars in $\omega$~Cen, thus providing a chemodynamical signature 
of their prior membership to this cluster. 

Why focus on HBA stars instead of, say, red HB/red clump stars, RR Lyrae stars, 
or RGB stars? First of all, HBA stars are almost exclusively associated 
with the Galactic halo, unlike red HB/red clump and RGB stars, which 
comprise a much more heterogeneous mix and whose properties are much more difficult 
to disentangle from one another. Also, while $\omega$~Cen possesses a vast population 
of HBA stars, it is almost devoid of red HB/red clump stars 
(see, e.g., Fig.~1 in \citealt{Ferraro2004}).
 Moreover, HBA stars being non-variable, their spectra are much easier 
to interpret 
than in the case of RR Lyrae variables. Last but not least, B-type blue HB (BHB) stars 
show strong abundance anomalies which do not reflect the abundances they had by the time 
they reached the RGB tip, being due instead to gravitational diffusion and radiative 
levitation (\citealt{Moehler2004} and references therein). 
 
HB stars are the immediate progeny of RGB tip stars. As such, one expects to 
find a close similarity between photospheric abundances for RGB tip stars and HBA 
stars belonging to the same population. This implies that field HBA stars 
which have once belonged to $\omega$~Cen should have unique signatures that would 
clearly set them apart from other HBA stars, due to the fact that 
RGB stars in $\omega$~Cen are well known to have peculiar abundance patterns 
in O, Na, Al, Mg, Cu, Eu, and the s-process elements, 
including extreme over- and 
underabundances compared to field stars with similar [Fe/H]
(some of the differences being themselves a function of [Fe/H]; 
e.g., \citealt{Norris1995_1,Smith2000,Pancino2002}). 
Such abundance patterns are not only unparallelled in other GCs; they have actually 
never been found among field halo stars (e.g., \citealt{Gratton2000}). 
In this sense, it would be extremely unlikely for a field star 
moving in a retrograde orbit with abundances 
$[{\rm O/Fe}] \le -0.4$, $[{\rm Al/Fe}] \ge +0.75$, $[{\rm Mg/Fe}] \ge +0.6$,
$[{\rm Cu/Fe}] \le -0.5$, $[{\rm Eu/Fe}] \le +0.1$, $[{\rm Ba/Fe}] \ge +0.75$, 
$[{\rm La/Fe}] \ge +0.75$   
{\em not} to have once been associated with $\omega$~Cen. Therefore, 
identifying HBA stars in the halo field with kinematics {\em and} detailed abundance 
signatures which are both consistent with former membership in the cluster would provide 
compelling evidence of the presence of $\omega$~Cen debris in the field.

\section{Sample composition and data}
\label{data}
Our sample consists of all known and unambiguously classified local ($d\le 1$~kpc)
HBA stars for which the 
required data, such as radial velocities and proper motions, are available.
We restricted ourselves to HBA stars, i.e. blue HB stars with 
a temperature of less than 10\,500~K, since hotter HB stars show strong peculiarities in 
their abundances due to radiative levitation and diffusion processes in their atmospheres
(\citealt{Moehler2004} and references therein), thus not allowing access to the envelope 
abundances of their immediate progenitors (RGB tip stars). 
We took all HBA stars from \citet[ hereafter AdB00]{AB2000},
and added the bona-fide HBA stars from \citet[ hereafter K00]{Kinman2000} and 
\citet[ hereafter B03]{Behr2003}. 
A total of 30 stars made it into our final sample, more than twice the 
number in AdB00. 
 
All the relevant data were taken from the literature. Most of the stars have Hipparcos 
\citep{HIP} parallaxes and proper motions, except in a few cases, where we had to rely 
on the slightly less accurate Tycho2 \citep{Tycho2}
proper motions. Since the parallaxes are mostly too inaccurate to directly derive the 
distances, we chose the approach used in AdB00, to which we refer the interested reader 
for further details. 

The radial velocities were also taken from AdB00,
K00 and B03. Comparison among the different sources shows a
generally good agreement, with the exception of HD 117880 where the value used in
AdB00 (taken from \citealt{Evans67}) differs by almost 200~\kms\
from the other values. Since our own low-resolution spectra show radial velocities rather 
similar to K00 and B03, we conclude that the velocity from Evans must be 
incorrect. For the errors in the kinematics we quote the values given in AdB00, of about 
10~\kms\ on average, with some dependence on distance. 

Our primary source for abundances, \teff, and \logg\ is K00. Not only is this study 
more complete than B03, it also
has more element species than just Mg and Fe. While for those stars which are included 
in both studies the general agreement for [Fe/H], \teff, \logg\ is quite good, there is the 
exception of
HD117880, for which the results of B03 differ significantly from those of K00
and other sources in the literature. K00 have [Mg/H] values for all of their programme stars,
[Fe/H] and [Ti/H] for most\footnote{In a few cases, [Fe/H] had to be derived from [Mg/H]. 
However, comparison between the K00 [Fe/H] values derived directly or via [Mg/H] does not 
reveal noticeable systematic discrepancies.} and [Ba/H] for a few. 
For some objects, we were able to include abundances of some additional elements from 
\citet{Adelmann1996} (and their earlier work).

The proper motions, radial velocities, distances, etc. were converted to
the $XYZ\,UVW\,\Theta\Phi$ coordinate system and 
orbits calculated using the potential of \citet{AS91}, 
as described in more detail in 
AdB00 and \citet{AEB2002}.
One of the quantities derived is $I_z$, the angular 
momentum in the $z$-direction,\footnote{Also denoted $J_z$ or $L_z$ in many publications.} 
representing the
orbital motion in the Galactic plane. In contrast to the related orbital velocity $\Theta$,
this is a conserved quantity --- and plays a crucial role in the following analysis. 
The velocity errors translate 
into an error of 80~kpc\,\kms\  for $I_z$ at the position of the Sun.

\begin{figure}

   \centering
   \epsfig{file=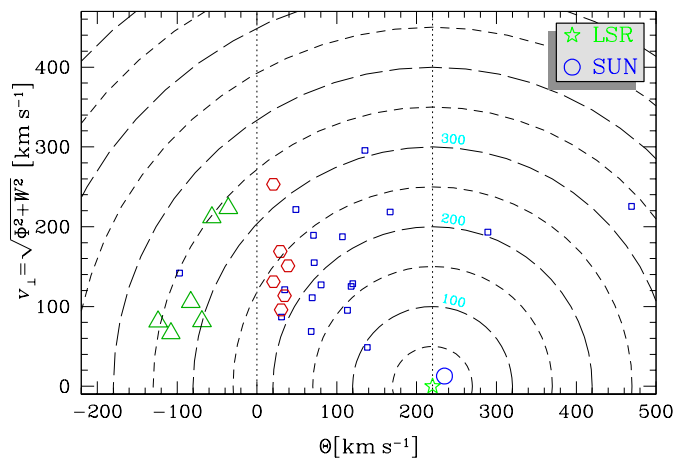,scale=0.45}
   \caption{Toomre diagram for the HBA stars in our sample. 
The concentric circles 
show the absolute peculiar velocity 
($v_{\rm pec}$), i.e. the
total deviation from circular velocity (shown as an open star; the
Sun's position in this diagram is shown as an open circle). 
The hexagons depict the stars belonging to the members of the 
Cometary Orbit Group (COG) discussed in this
paper, the triangles denote putative $\omega$ Cen debris candidates (identified
as retrograde moving stars broadly conforming to the abundance range of $\omega$~Cen giants),
the other HBAs are represented by squares.
 Note that the COG stars are relatively close together in this plane, their larger spread in 
$v_{\perp}$ most likely caused by their chaotic orbits. 
The most deviant point of this group is HD~86986. The rather large ($\ge$100~\kms) 
$v_{\rm pec}$ reveals that most probably all objects belong to the halo.
Stars of the thin disk have a $v_{\rm pec}$ of mostly less than 50~\kms, while 
some thick disk stars may have a $v_{\rm pec}$ of up to 150~\kms.}
   \label{toomre.fig}
\end{figure}

\begin{figure}

   \centering
   \epsfig{file=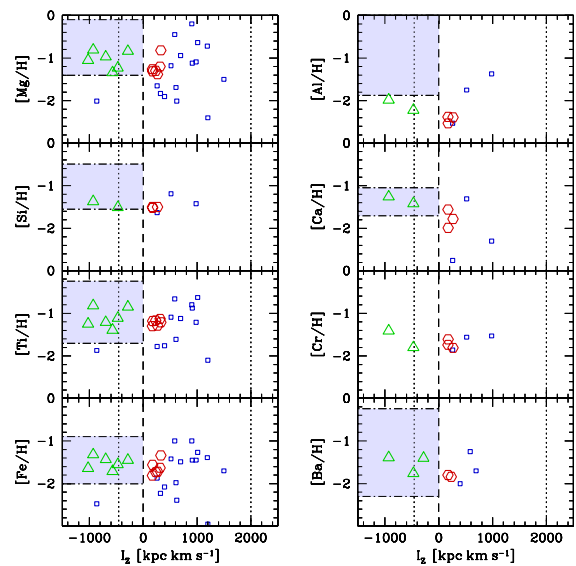,scale=0.40}
   \caption{[X/H] vs. $I_z$ diagram for eight different elements. 
    The vertical dashed line marks the boundary
    between pro- and retrograde rotation, while the dotted lines 
    $\omega$~Cen's $I_z$ (left) and that of the Sun (right) to represent the 
    $I_z$ of a typical disk star. The grey area between the horizontal lines in
    the retrograde part of the plots denotes the abundance
    range of each element in $\omega$~Cen, as conservatively derived from
    the extreme values of the relatively small sample of stars analysed by Smith et al. 
    (2000).
    The object symbols are used as described in Fig. \ref{toomre.fig}. 
    Two extreme objects are outside the $I_z$ range of this plot.
}
   \label{M_Iz.fig}
\end{figure}

\section{Results}
\label{results}
The new, enlarged sample confirms AdB00's basic results --- all HBA stars belonging to the halo. 
This can clearly be seen in the Toomre diagram shown in Fig. \ref{toomre.fig}, especially in 
comparison with similar diagrams in \citet{AEB2002} and \citet{Kaempf2005}.
One or two stars could in principle belong to the low-velocity 
end of the thick disk, but since thick disk stars with more ``normal'' 
orbital velocities are absent and given the fact that there are stars with rather similar but 
retrograde velocities, we conclude that we are still dealing with a pure halo sample, even more 
so since none of the potential thick disk candidates is particularly metal-rich. 

Two stars have orbital velocities significantly larger than $\Theta_{\rm LSR}$ (the disk velocity), 
one of which (HD 213468) with an extreme $\Theta = 470$~\kms\ and 
an apogalactic distance of over 200~kpc! We have thus found the first kinematic evidence 
of HBA stars belonging to the ``high velocity
halo,'' similar to the sdB stars found in \citet{AEB2002} and objects like Barnard's star.  
For this reason, the average orbital velocity is 58~\kms\,, compared to the 17~\kms\ found 
in AdB00. Removing these two stars results in $\bar{\Theta} \approx 35\,{\rm km}\,{\rm s}^{-1}$. 
However, the dispersion $\sigma_\Theta$ is much higher in this new sample than
the 55~\kms\ of AdB00, namely $80-120\,{\rm km}\,{\rm s}^{-1}$, 
depending on whether the two ``fast'' stars are included or not. 
Note that the distribution of orbital velocities shows signs 
of bimodality, with the prograde sample seemingly separated from the retrograde group by a gap. 
Furthermore, we found evidence of substructure in the kinematics of our HBA stars, which 
may be related to the halo's merger history.

Seven of our stars (23\%) have retrograde orbits, similar to $\omega$ Cen itself 
($\Theta = -69$~\kms, $I_z = -457$~kpc\,km\,s$^{-1}$). \citet{Bekki2003} and 
\citet{Tsuchiya2004} have calculated  
scenarios for the merger event forming the present-day $\omega$~Cen, predicting the kinematics 
of remnant stars in the solar vicinity. If these merger scenarios hold true,  
at least some of our retrograde stars could indeed represent debris from  
this merger event. However, as discussed in Sect.~2, a clear-cut signature of 
prior association can only be provided 
by a {\em very detailed} abundance analysis (see also \citealt{Dinescu2002}). 

\subsection{The Cometary Orbit moving group (COG)}
\label{res_prograde}
In the course of our search for $\omega$~Cen candidates, 
we serendipitously found a very striking group with common properties among our 
{\em prograde} stars. This group, consisting of four to six stars,\footnote{HD~106304 
has abundances which are slightly discrepant with respect to the other COG stars and HD 86986 has somewhat different radial kinematics, see Fig. \ref{toomre.fig}.}  
is located near $I_z=+230$~kpc\,\kms and ${\rm [Fe/H]}=-1.7$, according to Fig.~\ref{M_Iz.fig}.
In this figure, we plot the abundances of several different species as a function of $I_z$. 
The COG is very confined in $I_z$ ($\sigma_{I_z}=55$~kpc\,\kms) and in most element species.
The exception is Ca; however, we only have Ca values for three stars, from \citet{Adelmann1996}.

All stars in this group have orbits taking them to less than 1~kpc of the Galactic centre --- in 
some cases, to less than 300~pc.
They are on chaotic orbits, which means that inclination and maximum distance from the plane changes with
every revolution, and even small inaccuracies in the input values have a large effect on the shape (but
not the overall size) of the orbit. Therefore the tangential velocities, $z_{\rm max}$, $nze$, etc. do not
really help in clarifying whether we are really dealing with a distinct group of stars; the eccentricity
is over 0.9 for all stars, given their small perigalactic distances. Nonetheless, the evidence for this
group is rather strong since we have these groupings in 7 of 8 element species, and especially in those
where we have data for (nearly) all stars (Mg, Fe, Ti). 
In the near future we will be able to shed more light 
on the issue, when we have a consistent set of abundances for many elements. 

A remarkable characteristic 
of the COG is that it goes so near the Galactic centre. Does 
this mean that it partook the initial collapse of the halo, as described in \citet{ELS62}? 
Does its small spread in chemical abundances indicate instead that 
it is a disrupted 
GC which was formerly on a radial orbit? Alternatively, this group could be the remains
of yet another merging event, 
maybe one that took place in the relatively remote past (i.e., at 
least one revolution ago), and whose orbits have since
decayed to the current cometary form. In the data of \citet{Peterson2001},  
who studied HBA stars in a window towards the 
bulge, we found a slight overdensity of stars with  
${\rm [Fe/H]} \sim -1.5$ (their values are only good 
to 0.5~dex) at a line of sight velocity of 100~\kms\ \citep{Altmann2002}.
In principle, these could be related to the stars discussed here; we would need the full 
kinematics (including accurate proper motions) to resolve this issue. 
This adds, however, to the evidence of inhomogeneities in the (inner) halo.

\section{Summary and outlook}
\label{outlook}
In this {\em Letter}, we have reported on the discovery of a group (the COG) of (local) HBA 
stars with similar (prograde) orbits and abundances, which may represent the remains of a 
protogalactic fragment. Their cometary orbits take the stars very close to the Galactic 
centre. The very small spread 
in abundances among the COG stars may alternatively support the possibility that they come 
from a disrupted GC. This group was discovered while we were testing a new method to search for 
debris of specific merger events associated with identified surviving remnants, such as 
$\omega$~Cen. 
The method uses orbital information 
and very detailed, multi-species abundance patterns to search for the chemical signatures of 
surviving merger remnants among (local) field stars having the appropriate kinematics. The 
rationale of the method is that a few key elements, such as O, Na, Mg, Al, Cu, Eu and the 
s-process elements, may show unique signatures in the spotted surviving fragment (as has 
been seen in the case of $\omega$~Cen in particular), so that identifying field stars with similar 
abundances should represent the ``smoking gun'' indicating the presence of related tidal debris 
in the field. 

More detailed abundances will both shed light into the case of a possible $\omega$~Cen remnant and
further establish the reality and nature of the group dealt with in this {\em Letter}.

\begin{acknowledgements}
      MA and MZ are supported by Fondap Center for Astrophysics 15010003, 
      and MC by Proyecto FONDECYT Regular 1030954. 
      We warmly thank M. Geffert and M. Odenkirchen for the 
      kinematic software they readily supplied to us,  
      and D. Dinescu, I. Ivans, B. J. Pritzl, and an anonymous referee for 
      useful comments and discussions. 
      With pleasure we made extensive use of the SIMBAD archive at CDS.
\end{acknowledgements}

\bibliographystyle{aa}
\bibliography{Hb014}

\begin{thebibliography}{28}
\expandafter\ifx\csname natexlab\endcsname\relax\def\natexlab#1{#1}\fi

\bibitem[{{Adelman} \& {Philip}(1996)}]{Adelmann1996}
{Adelman}, S.~J. \& {Philip}, A.~G.~D. 1996, MNRAS, 280, 285

\bibitem[{{Allen} \& {Santillan}(1991)}]{AS91}
{Allen}, C. \& {Santillan}, A. 1991, RMxA, 22, 255

\bibitem[{{Altmann}(2002)}]{Altmann2002}
{Altmann}, M. 2002, {PhD~Thesis; Kinematics and Population Membership of BHB
  and EHB Stars} (Sternwarte der Univ. Bonn)

\bibitem[{{Altmann} \& {de Boer}(2000)}]{AB2000}
{Altmann}, M. \& {de Boer}, K.~S. 2000, A\&A, 353, 135

\bibitem[{{Altmann} {et~al.}(2004){Altmann}, {Edelmann}, \& {de
  Boer}}]{AEB2002}
{Altmann}, M., {Edelmann}, H., \& {de Boer}, K.~S. 2004, A\&A, 414, 181

\bibitem[{{Behr}(2003)}]{Behr2003}
{Behr}, B.~B. 2003, ApJS, 149, 101

\bibitem[{{Bekki} \& {Freeman}(2003)}]{Bekki2003}
{Bekki}, K. \& {Freeman}, K.~C. 2003, MNRAS, 346, L11

\bibitem[{{Dinescu}(2002)}]{Dinescu2002}
{Dinescu}, D.~I. 2002, in ASP Conf. Ser. 265: Omega Centauri, A Unique Window
  into Astrophysics, 365

\bibitem[{{Eggen} {et~al.}(1962){Eggen}, {Lynden-Bell}, \& {Sandage}}]{ELS62}
{Eggen}, O.~J., {Lynden-Bell}, D., \& {Sandage}, A.~R. 1962, ApJ, 136, 748

\bibitem[{ESA(1997)}]{HIP}
ESA. 1997, The Tycho and Hipparcos catalogue, Vol. SP-1200 (ESA)

\bibitem[{{Evans}(1967)}]{Evans67}
{Evans}, D.~S. 1967, in IAU Symp.: Determination of Radial Velocities and their
  Applications, Vol.~30, 57

\bibitem[{{Ferraro} {et~al.}(2004){Ferraro}, {Sollima}, {Pancino},
  {Bellazzini}, {Straniero}, {Origlia}, \& {Cool}}]{Ferraro2004}
{Ferraro}, F.~R., {Sollima}, A., {Pancino}, E., {et~al.} 2004, ApJ, 603, L81

\bibitem[{{Gratton} {et~al.}(2000){Gratton}, {Sneden}, {Carretta}, \&
  {Bragaglia}}]{Gratton2000}
{Gratton}, R.~G., {Sneden}, C., {Carretta}, E., \& {Bragaglia}, A. 2000, A\&A,
  354, 169

\bibitem[{{Hilker} {et~al.}(2004){Hilker}, {Kayser}, {Richtler}, \&
  {Willemsen}}]{Hilker2004}
{Hilker}, M., {Kayser}, A., {Richtler}, T., \& {Willemsen}, P. 2004, A\&A, 422,
  L9

\bibitem[{{H{\o}g} {et~al.}(2000){H{\o}g}, {Fabricius}, {Makarov}, {Urban},
  {Corbin}, {Wycoff}, {Bastian}, {Schwekendiek}, \& {Wicenec}}]{Tycho2}
{H{\o}g}, E., {Fabricius}, C., {Makarov}, V.~V., {et~al.} 2000, A\&A, 355, L27

\bibitem[{{Ibata} {et~al.}(1994){Ibata}, {Gilmore}, \& {Irwin}}]{Ibata94}
{Ibata}, R.~A., {Gilmore}, G., \& {Irwin}, M.~J. 1994, Nature, 370, 194

\bibitem[{{Kaempf} {et~al.}(2005){Kaempf}, {de Boer}, \&
  {Altmann}}]{Kaempf2005}
{Kaempf}, T.~A., {de Boer}, K.~S., \& {Altmann}, M. 2005, A\&A, 432, 879

\bibitem[{{Kinman} {et~al.}(2000){Kinman}, {Castelli}, {Cacciari}, {Bragaglia},
  {Harmer}, \& {Valdes}}]{Kinman2000}
{Kinman}, T., {Castelli}, F., {Cacciari}, C., {et~al.} 2000, A\&A, 364, 102

\bibitem[{{Layden} \& {Sarajedini}(2000)}]{Layden2000}
{Layden}, A.~C. \& {Sarajedini}, A. 2000, AJ, 119, 1760

\bibitem[{{Martin} {et~al.}(2004){Martin}, {Ibata}, {Bellazzini}, {Irwin},
  {Lewis}, \& {Dehnen}}]{Martin2004}
{Martin}, N.~F., {Ibata}, R.~A., {Bellazzini}, M., {et~al.} 2004, MNRAS, 348,
  12

\bibitem[{{Mizutani} {et~al.}(2003){Mizutani}, {Chiba}, \&
  {Sakamoto}}]{Mizutani2003}
{Mizutani}, A., {Chiba}, M., \& {Sakamoto}, T. 2003, ApJ, 589, L89

\bibitem[{{Moehler}(2004)}]{Moehler2004}
{Moehler}, S. 2004, in The A-Star Puzzle, IAU Symp. 224, ed. J. Zverko, J.
  Ziznovsky, S. J. Adelman, \& W. W. Weiss (San Francisco: IAU), 395

\bibitem[{{Norris} \& {Bessel}(1975)}]{Norris1975}
{Norris}, J. \& {Bessel}, M.~S. 1975, ApJ, 201, L75

\bibitem[{{Norris} \& {Da Costa}(1995)}]{Norris1995_1}
{Norris}, J.~E. \& {Da Costa}, G.~S. 1995, ApJ, 447, 680

\bibitem[{{Pancino} {et~al.}(2002){Pancino}, {Pasquini}, {Hill}, {Ferraro}, \&
  {Bellazzini}}]{Pancino2002}
{Pancino}, E., {Pasquini}, L., {Hill}, V., {Ferraro}, F.~R., \& {Bellazzini},
  M. 2002, ApJ, 568, L101

\bibitem[{{Peterson} {et~al.}(2001){Peterson}, {Terndrup}, {Sadler}, \&
  {Walker}}]{Peterson2001}
{Peterson}, R.~C., {Terndrup}, D.~M., {Sadler}, E.~M., \& {Walker}, A.~R. 2001,
  ApJ, 547, 240

\bibitem[{{Smith} {et~al.}(2000){Smith}, {Suntzeff}, {Cunha}, {Gallino},
  {Busso}, {Lambert}, \& {Straniero}}]{Smith2000}
{Smith}, V.~V., {Suntzeff}, N.~B., {Cunha}, K., {et~al.} 2000, AJ, 119, 1239

\bibitem[{{Tsuchiya} {et~al.}(2004){Tsuchiya}, {Korchagin}, \&
  {Dinescu}}]{Tsuchiya2004}
{Tsuchiya}, T., {Korchagin}, V.~I., \& {Dinescu}, D.~I. 2004, MNRAS, 350, 1141

\end{thebibliography}

\end{document}